# Software Defined Networking: An Overview

| Alexander Nunez | Joseph Ayoka | Md Zahidul Islam | Pablo Ruiz |
|---|---|---|---|
| *Dept. of Electrical and Computer Engineering* | *Dept. of Electrical and Computer Engineering* | *Dept. of Electrical and Computer Engineering* | *Dept. of Electrical and Computer Engineering* |
| *University of Massachusetts* | *University of Massachusetts* | *University of Massachusetts* | *University of Massachusetts* |
| Lowell, MA 01854, USA | Lowell, MA 01854, USA | Lowell, MA 01854, USA | Lowell, MA 01854, USA |

*Abstract*— The Internet is the driving force of the new digital world, which has created a revolution. With the concept of the Internet of Things (IoT), almost everything is being connected to the internet. However, with the traditional IP network system, it is computationally very complex and costly to manage and configure the network, where the data plane and the control plane are tightly coupled. In order to simplify the network management tasks, software-defined networking (SDN) has been proposed as a promising paradigm shift towards an externalized and logically centralized network control plane. SDN decouples the control plane and the data plane and provides programmability to configure the network. To address the overwhelming advancement of this new technology, a holistic overview of SDN is provided in this paper by describing different layers and their functionalities in SDN. The paper presents a simple but effective overview of SDN, which will pave the way for the readers to understand this new technology and contribute to this field.

## I. INTRODUCTION

IN a traditional IP network, the network devices communicate with each other through various defined protocols to negotiate the exact network behavior based on the configuration of each device. In this architecture, there is no significant process of automation, and the network administrator is required to manually configure a large number of devices. Though some software tools facilitate the work, it still requires a significant amount of time and human resources. An example of a simple software tool is the simple network management protocol (SNMP), which is widely used for statistics and alarm collection but not for configuration management due to numerous limitations as listed in [1]. There is another protocol, NETCONF, to automate the configuration of network devices through a standard application programming interface (API). However, before using the NETCONF, all of the functionalities and logic must be implemented on the network devices [1].

Another problem with traditional networking is that the network devices are sold as closed components at a very high price by the manufacturers with their own operating systems, specific hardware, and features. As there are many equipment vendors with their own configuration languages to meet the network protocols, there can be several interoperability issues for the integration of devices from different vendors. This closed component also hinders the participation of diverse research communities to bring innovation to computer networks. Furthermore, the traditional network is not adaptable when it comes to adding new services to an existing network. A new device must be integrated and configured to perform the new service before it can be added [2].

In contrast to the traditional networking, software-defined networking (SDN) has introduced a paradigm shift in computer networking by decoupling the control plane from the data plane. The switches and the routers become pieces of hardware in the data plane to redirect the traffic from one interface to another, whereas the controller in the control plane communicates with the switches to control the switches based on user-defined protocols. As all the switches can be managed from the central controllers, it is no longer required to access individual switches and configure them. The network controller communicates with the switches using standard protocols like OpenFlow, so there is no need to use vendor-specific programs to control the forwarding behavior of the switches, which facilitates the interoperability between devices from different vendors. [2].

SDN also provides an easy interface for implementing new services in the network. Any service in the network can be implemented by writing the logic in a high-level programming language and having the controller translate this logic into low-level forwarding rules for the switches. SDN shifts the focus from hardware to software and provides the opportunity for the diverse research communities to develop their own ideas and implement them.

The details of SDN structure and functionalities are described in this paper. The definition of SDN is described in Section II; the data plane and the control plane of SDN are discussed in Sections III and IV, respectively. Sections V and VI summarize the management plane and challenges of SDN, respectively, and finally, Section VII concludes the paper.

## II. SOFTWARE DEFINED NETWORKING

### A. What is SDN

A Software Defined Network (SDN) is a networking approach that vastly differs from the internet's traditional network structure. Software-defined networking is a networking method that uses software-based controllers and application programming interfaces (APIs) to communicate with underlying hardware infrastructure (switches and routers) and to direct traffic on a network. With the use of a centralized software-based SDN controller, the network can be controlled and automated via software. Software-defined networking allows for vertical network integration, separation of the network's control logic from the underlying routers and switches, centralized logical network control, and the ability to program the network [3]. To further understand what SDN is, it is important to know its architecture. The SDN architecture has 3 planes: the management plane, the control plane (SDN controller operates here), and the data plane (switches, forwarding devices) [3]. SDN provides centralized network control by separating the control plane and the data plane from both being on routers. Instead of having the control plane governed by protocols on routers and switches, the control plane is now managed by the SDN controller. The controller is connected to all the underlying switches and routers found in the data plane. This allows the controller to control the network devices by dispatching different commands and instructions to them. Software-defined networking can be thought of as separating the "brains" from the power in a network. The controller would be the "brains" in this example because it has the management capabilities and instructions to control the power (routers and switches).

*B. How does SDN Work*

Now that it is clear what software-defined networking is, it is important to learn an overview of how it works. As stated before, SDN detaches the control plane and the data plane from being on a single device like a router. The detached control plane is now known as the SDN controller. Having a main controller gives network administrators the ability to control the entire network via one controller instead of on a device by device basis. An SDN network allows for dynamic, efficient, and centralized network configuration. Dynamic network configuration is possible through the use of SDN applications. SDN applications are used to configure the network and are created with specified network requirements and desired network behavior. Applications are in the management plane and they are deployed by the SDN controller to the switches and routers in the data plane. The applications allow SDN controllers to autonomously control the network and its behavior. In order for this to be possible there needs to be a way for the SDN controller to communicate with every device on the network. The 3 SDN planes use dedicated interfaces to interact with each other. There are two interfaces named the northbound and southbound interfaces. The northbound interface connects the SDN controller/control plane to the management plane which is where applications are deployed and handled. While the southbound interface connects the controller to the underlying forwarding devices. The southbound interface allows for control of the devices, forwarding operations, and other services [3]. Communication through these interfaces is possible with the use of an Application Programming Interface (API).

*C. Motivation behind SDN*

Why should software defined networks be implemented? An example can be if there is a demanding application that a traditional network cannot handle or deploy without manual and costly configurations. Also, if there is a need for centralized control of a network to quickly adjust to bottlenecks and application requirements. Centralized control over a network is crucial in the cloud because it's necessary to move data between distributed systems and SDN allows data to move easily between distributed locations. All these advantages are possible with software defined networking, that is otherwise difficult to configure onto a traditional network. Some drawbacks traditional networks have that SDN helps address include: moving from distributed network architecture to centralized network architecture, from a non-programmable network to a programmable and configurable network, and from static/manual configuration of devices to autonomous configuration. SDN also has special services that supply useful network data and statistics. SDN provides the following services [3]:

- Topology service (Determines how forwarding devices are connected to each other and assembles a topology of the network)
- Inventory service (used to record all SDN devices and basic information about them (Ex: The version of OpenFlow/APIs used and its capabilities).
- Statistics service (flow tables and flow counters)
- Host tracking (determines where IP and MAC addresses of hosts are located on the network)

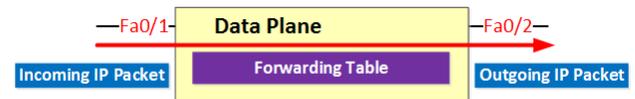

Fig. 1. Data Plane Representation [4].

## III. DATA PLANE OF SDN

*A. Overview of the Data Plane*

The function of the data plane in a SDN controlled network does not change fundamentally compared to a traditional network with a distributed control plane [4]. The main purpose of the data plane is to forward arriving datagrams from their respective input port to the correct output port as determined by the forwarding or flow tables. The critical performance metric for the data plane is data throughput, or in other words how many datagrams or bytes can it forward per unit time. To maximize the performance of the data plane in this metric it is critical that it be implemented using custom hardware at each network router or switch. This specialized hardware commonly includes ternary content addressable memories (TCAMs) which are very fast content addressable tables which can retrieve the address of a field in one clock cycle. A good example is the Cisco Catalyst network switch product line which includes TCAMs with approximately one million forwarding table entries [5].

*B. Forwarding Logic*

In a traditional distributed network, the data plane relies on forwarding tables which are populated by the control plane to tell it what output port each datagram should be forwarded to. Forwarding tables are relatively simple containing only two columns, the first of which contains a destination address range and the second of which contains the corresponding link interface or output port. The destination address range fields in the forwarding table are matched to the address obtained from each datagram's destination address header field. In the case where there are one or more ranges which match the destination address the longest will be selected as it represents a match to a more specific, less broad subnet.

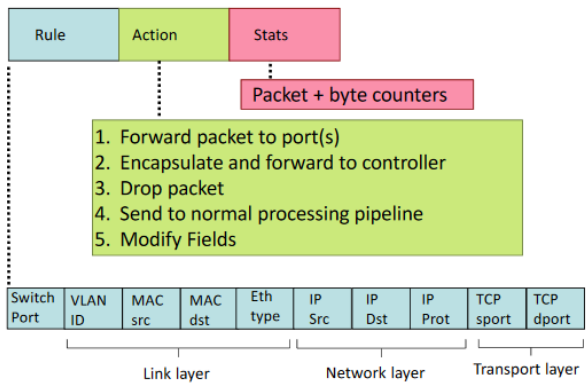

Fig. 2. Flow Table Entry Representation [6].

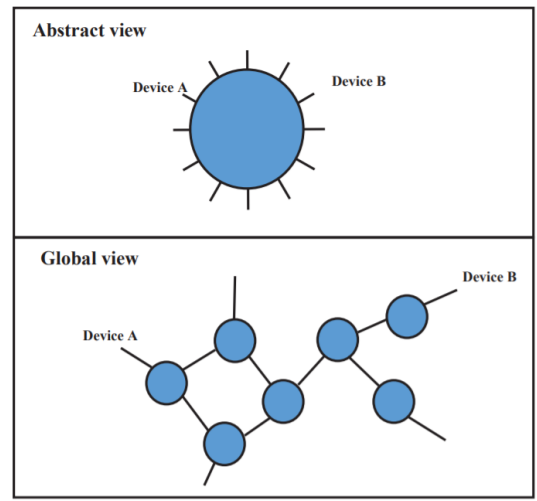

Fig. 3. Global and Abstract view of Network [2].

On the other hand, SDN data planes use an improved, more complex version of forwarding tables referred to as flow tables. These new tables allow the data plane to perform more complex decisions based on any of the data contained within a datagram. Flow tables contain one more column compared to forwarding tables bringing the total up to three columns. The first column fields contain the rules for matching incoming packets to specific actions. These rules can be applied to any part of the incoming datagram including its data payload unlike the rules in forwarding tables which can only be applied to the destination address field in each datagram's header. The fields in the second column of a flow table contain the actions to be performed on an incoming datagram that matched with the corresponding rule field. These actions could be virtually anything within the capabilities of the data plane hardware, from forwarding a datagram to an output port to modifying fields within the datagram and even outright dropping the packet. The fields in the third and final column of a flow table are used to store performance metrics on their corresponding rule and action fields. For example, they are commonly used to count the number of datagrams that triggered the corresponding rule, or the number of bytes contained within said packages.

Together the three fields within flow tables offer vastly improved functionality over a simple forwarding table. However, this functionality comes at the cost of increased performance overheads and increased complexity within the network's core. These downsides would have made flow tables impractical in the past but the continuous increases in compute performance and efficiency have made it both viable and economical [4].

### B. OpenFlow Protocol

Since in SDN the control plane and data plane are not physically on the same device a new software interface had to be developed as a means of communication. There are a variety of proprietary interfaces of this kind however the most common is an open-source protocol named OpenFlow which is maintained by the Open Network Foundation [7].

The OpenFlow protocol lays out the guidelines for communication between a "dumb" network device which only contains the data plane and the centralized control plane. It uses the TCP networking protocol to establish a connection with network devices and transmits updated flow table entries to the device from the centralized control plane. OpenFlow is strictly a protocol for communication between the control plane and the data plane, it is not used for communication between data plane devices [7].

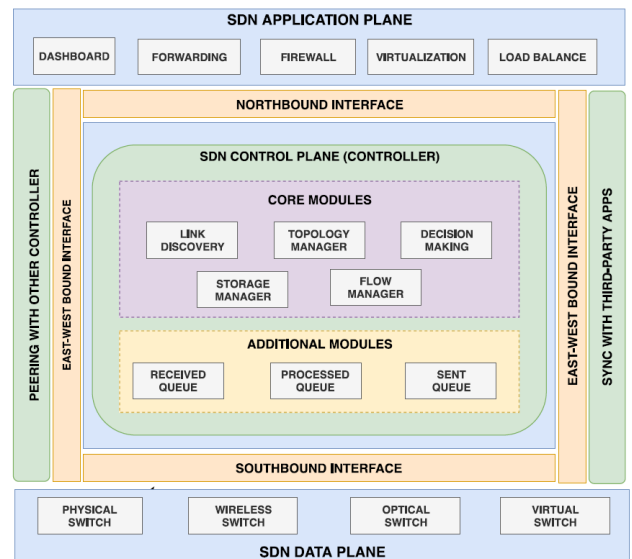

Fig. 4. Overview of SDN controller [8].

## IV. CONTROL PLANE OF SDN

### A. Overview of the Control Plane

The control plane is separated from the data plane and communicates and manages the data plane using various protocols like OpenFlow. It is responsible for creating the routing table of the switches and managing the traffic in the switches. The control plane consists of two components, which are the SDN controllers and a set of network-control applications. The controller is the core component in SDN infrastructure, which is responsible for managing and controlling the controlled switches. The basic structure of a SDN controller and its functionality is described below.

*1. Controller Core:* The controller core interacts with the controlled switches, maintains the global view of the underlined networks, and creates an abstract view for the application layer.

| | NOX | POX | Beacon | Ryu | Floodlight | ODL | ONOS |
|---|---|---|---|---|---|---|---|
| First Release | 2008 | 2011 | 2010 | 2012 | 2012 | 2013 | 2014 |
| License | GPL 3.0 | Apache 2.0 | GPL 2.0 | Apache 2.0 | Apache 2.0 | EPL 1.0 | Apache 2.0 |
| Multi-Thread | No | No | Yes | Yes | Yes | Yes | Yes |
| Platform | Linux | Linux MAC windows | Linux MAC windows | Linux MAC windows | Linux MAC windows | Linux MAC windows | Linux MAC windows |
| APIs | OVSDB OFREST | OVSDBOF REST | OVSDB OF REST | OVSDB, OF, REST NETCONFO FCONFIG | OVSDBOF REST | OVSDB,OFREST,YA NG NETCONFPCEP, BGPSNMP | OVSDB, OF REST, NETCONF |
| OF Version | 1 | 1 | 1 | 1.0 to 1.5 | 1.0 to 1.3 | 1.0 to 1.3 | 1.0 to 1.3 |
| Documentation | poor | poor | poor | Medium | Very Good | Very good | Medium |
| Language | C++ | python | Java | Python | Java | Java | Java |
| Legacy Network | No | No | No | No | No | Yes | Yes |
| Category | Single | Single | Single | Single | Single | Distributed | Distributed |
| GUI | Poor | Poor | Poor | Medium | Medium | Very Good | Very Good |
| Regular Updating | Poor | Poor | Poor | Medium | Medium | Very Good | Very Good |
| Modularity | Poor | Poor | Medium | Medium | Medium | Very Good | Very Good |

Fig. 5. Features of different SDN controllers [9].

For example, as shown in Fig. 3, several switches are necessary in order to communicate between Device A and Device B, which is marked as Global view. The controller obtains this global view of the network and create the abstract view of the network. The abstract view over the network is accessed to write a control program by the user. There can be several modules for doing specific tasks in controller as shown in Fig. 4. The link discovery modules regularly interact with the switches to build the topology of the network, which is then maintained in the topology manager. The other modules such as decision making module use the network topology from the topology manager and find optimal paths between nodes in the network. The flow modules directly interact with the flow tables of the data plane to notify them the updated tables based on the rules. In addition, the controller may have storage and queue manager for storing various information and queue reports.

*2. Interfaces:* The core controller is surrounded by different APIs for interacting the different layers. The upper layer application uses northbound API to interact with the core, while the core uses southbound API to interact with the controlled switches. The east-westbound APIs can be used for interacting with controller themselves in case of distributed multiple controllers. The widely used SBI is OpenFlow protocol, however, with the recent controllers, the SBI supports other protocol along with the OpenFlow. In case of NBI, controllers support a number of protocols, but the most popular one is the REST API.

### B. Different Types of SDN Controllers

The most well-known SDN controllers are NOX, POX, Beacon, Floodlight, Ryu, OpenDayLight (ODL), and ONOS [10]. The different controllers have different and diverse features. In Fig. 5, a qualitative features of the controllers such as first release date, running platform, supporting APIs, programming language, network category, recent updating, etc., are shown. Among the controllers, the ODL and ONOS controllers are the most recent distributed open-sourced controllers. These controllers have their unique architecture based on their goals, however, the overview of their architecture can be generalized as shown in Fig. 4. A QoS based evaluation of ODL and ONOS is performed in [11]. For evaluating their performances, three different topologies, namely single, linear, and tree topologies are constructed with 8 hosts, 3 OpenFlow switches and 1 controller using Mininet [11]. The data transfer delay between two hosts is measured for transferring a large number of fixed size packets with certain inter-departure time. The time-delay and packet drops experienced by the ODL and ONOS controllers are summarized in Table I. It is observed that the ONOS controllers over-performs the ODL controller in all the three topologies.

Table I: Comparison between ODL and ONOS controllers [11].

| Topology | Controller | Max delay (ms) | Avg. delay (ms) | Packet dropped (%) |
|---|---|---|---|---|
| Single | ODL | 41.64 | 1.14 | 0.25 |
| | ONOS | 2.38 | 0.014 | 0 |
| Linear | ODL | 33.23 | 0.87 | 0.09 |
| | ONOS | 4.03 | 0.02 | 0.02 |
| Tree | ODL | 49.96 | 1.94 | 0.63 |
| | ONOS | 1.21 | 0.01 | 0 |

### C. Challenges in the Control Layer

*1. Controller Placement Problem:* Due to the decoupling of the data and control plane in SDN, the controlled-switches are memoryless and query to the controllers for appropriate flow table for every new flow it encounters. The controller processes the queries, sets up the required flow table, and then, send the flow tables to the switches. Therefore, in order to process the flow tables with minimum latency, the relative location between controller-switch needs to be optimized. Randomly placed controllers increase the latency in the network, which in turn creates a wide range of network issues such as resiliency, energy efficiency, load balancing, quality-of-service (QoS), and so on. Due to the importance of optimal controller placement (OCP) problem, it has been extensively studied in the literature. OCP problem can be addressed in several categories, such as how many controllers to place in an SDN network and where to place them in a network to achieve particular objectives. Some important objectives are network responsiveness, fault tolerance, resilience, QoS, load balancing, latency, and so on. A variety of methodologies based on graph theory, clustering algorithms, game theory, greedy algorithms, approximation algorithms, exhaustive search, etc., are described to attain the objectives in OCP [10]. The OCP problem is still an open problem in SDN and both the industries and the research communities are searching for an appropriate solution to meet the future requirements of SDN.

*2. Centralized vs. Distributed Controller:* The placement of SDN controllers can be categorized into two categories, namely centralized and distributed controller [12]. In centralized controller, only a single controller manages the whole network. This architecture works very well in an enterprise network in terms of better performance and user experiences. However, as user traffics increase in large networks such as data center, various issues related to security, scalability, performance, and availability arises. To alleviate these issues of centralized controller, distributed controllers are used, where multiple controllers are used to manage and obtain the global view of the network. The distributed controllers are divided into flat and hierarchical architecture. In the flat architecture, the whole network is split into small networks and a single controller mange the small network. On the other hand, in hierarchical architecture, multiple local controllers are used to manage local network and they provide information to the upper layer of controllers. In [12], the authors compare the performances of the two categories by using load balancing application. The results show that distributed controller is better in terms of response time and number of transactions. It also demonstrates that the distributed controllers are better for scalability and reduces the bottleneck in the network.

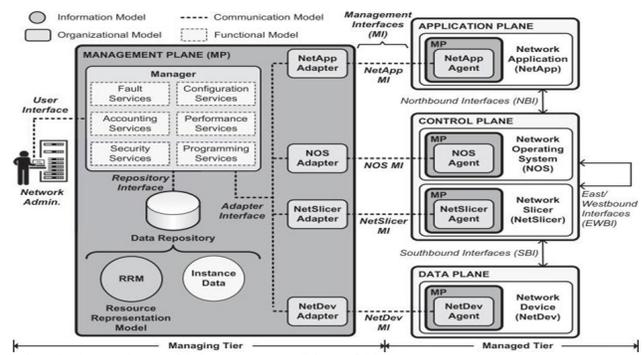
Fig. 6. Overview of Management Plane [13].

Repository Interface: This interface connects the Manager with the Data Repository and instance data.

Adapter Interface and Management Interface: These interfaces are responsible for the transportation of requested messages /details from the Manager or Network Admin to a particular Agent in a given plane. This is being channeled through the needed Adapter at a given time.

The other Management interfaces are shown is Fig. 6 as, NetApp MI, NOS MI, NetSlicer MI, and NetDev MI. These Management Interfaces handle data formats and protocols implemented as (*e.g.* NCP, SNMP, SSH) etc.

With these services at the hand of the Network Admin, SDN resources can be managed and controlled in order to achieve smooth performance.

## V. MANAGEMENT PLANE OF SDN

The Management plane is often referenced on its own and differs from the other planes in SDN. It is responsible for the Management and control of other planes. The management plane will be described with the help of Fig. 6, which shows how the management plane works and operates. The plane starts with the basic 5 management functional areas which are the Fault Services, Configuration Services, Accounting Services, Performance Services, and Security Services. They will be explained below.

Fault Services: The aim of Fault Services is to detect, separate, and correct failures which may be present in logical or physical Software Defined Networking resources.

Configuration Services: This modifies and updates the state of SDN resources.

Accounting Services: Accounting services does the job of constantly tracking and allocating usage of Software Defined Network resources.

Performance Services: It collects and reports information about the operation of SDN resources to the Network Admin.

Security Services: Security services is present in order to control and monitor access to SDN resources.

Programming Services: This service is present in order to assist in updating any programmable software of SDN resources.

As seen in Fig. 6, there are various interfaces in the management plane, they include the User Interface, Repository Interface, Adapter Interface, and the set of Management Interfaces. They are explained below.

User Interface: The user interface allows the network admins to interact with Management Services.

## VI. CHALLENGES IN SDN

Due to the centralized nature of SDN, its implementation can have beneficial effects on network security [14] By continually collecting data and network statistics from all the network switches within a network a centralized controller has a comprehensive view of the network performance and state. This unique position could allow it to detect network threats before they have a chance to compromise the network's performance [15].

For example, if a network threat such as a Distributed Denial of Service (DDoS) attack is detected the control plane could algorithmically find similarities between the malicious datagrams and then proceed to modify the flow tables of the affected routers, using the OpenFlow protocol, to identify and drop the targeted packets without affecting the rest of the network traffic [15].

Another, novel SDN security strategy is Moving Target Defense (MTD) algorithms. These algorithms aim to periodically change the location of key network features to make it more difficult for attackers to discern the composition of the network and the addresses of its critical components [16]. This strategy would be very hard and complex to implement in a traditional network but is relatively straight forward in an SDN due to the flexibility provided by having a central controller and having the rest of the network devices be interchangeable "dumb" switches.

## VII. CONCLUSION

In conclusion, implementing an SDN network mainly consists of separating the control plane from the data plane and then combining every router's control plane into a central entity that can define the overall network behavior as well as each individual switch's flow tables. Such a centralized control plane is defined primarily in software, and as such, it can interact directly with applications through its northbound interface, or API, which makes the network much more flexible and adaptable to each specific use case. Additionally, since the control plane is now separate from the data plane, it uses a southbound software interface, commonly OpenFlow, to communicate and populate the flow tables of each individual network switch. The above-mentioned features enable SDNs to quickly adapt to changing network requirements and performance.

Although SDN offers a multitude of improvements over traditional networking, it also comes with its own unique challenges. Due to the network's reliance on a centralized controller, it is at risk of collapsing if the controller is compromised. For this reason, backup, physically distributed controllers must be maintained to ensure network resiliency. Additionally, the controllers must be scalable to control and manage the ever-increasing number of network-enabled devices and their traffic. Finally, the new open interfaces of a SDN may introduce new forms of network attacks that need to be addressed in the SDN framework and communication protocols.

Overall, SDNs are a recent innovation and, as such, constitute a relatively small portion of the networks currently in use. However, the rise of distributed applications with more decentralized and complex traffic patterns, such as cloud computing and cloud services, which require more flexible networks, almost guarantees that SDNs will become more widespread going forward.